\documentclass[parskip=half]{scrartcl}

\usepackage[english]{babel}
\usepackage[T1]{fontenc}
\usepackage[utf8]{inputenc}

\usepackage{authblk} 

\usepackage{chngcntr}		
	\counterwithout{equation}{section}			

\usepackage{hyperref}   

\makeatletter
\renewcommand{\subsection}{\scr@startsection%
    {subsection}
    {2}
    {0mm}
    {0.5\baselineskip}
    {0.1\baselineskip}
    {\large\sffamily\bfseries}}
\renewcommand{\subsubsection}{\scr@startsection%
    {subsubsection}
    {3}
    {0mm}
    {0.5\baselineskip}
    {0.1\baselineskip}
    {\sffamily\bfseries}}
\makeatother

\usepackage{currfile}   

\usepackage[shortlabels]{enumitem}

\usepackage[labelformat=simple]{subcaption}
\usepackage{amssymb}
\usepackage{amsmath}
\usepackage{mathtools}
    \mathtoolsset{showonlyrefs=true}

\usepackage{amsthm}
    \newtheorem{thm}{Theorem}
    \newtheorem{lem}[thm]{Lemma}

    \theoremstyle{definition}
    \newtheorem{defi}[thm]{Definition}
    
\usepackage{bm}         
         
\usepackage[dvipsnames, table, xcdraw]{xcolor} 

\usepackage{orcidlink}
        
\newcommand{\doi}[1]{DOI \href{https://doi.org/#1}{#1}}

\newcommand{\refFig}[1]{Figure~\ref{fig:#1}}

\newcommand{\refSec}[1]{Section~\ref{sec:#1}}
\newcommand{\refLem}[1]{Lemma~\ref{lem:#1}}

\newcommand{\refThm}[1]{Theorem~\ref{thm:#1}}

\newcommand{\N}{\mathbb{N}}

\newcommand{\pause}{}   

\edef\path{\currfiledir}

\newcommand{\inputFromHere}[1]{%
    \edef\pathhere{\currfiledir}%
    
    \input{\pathhere/#1}

}

\usepackage{currfile}       

\usepackage{xifthen}        
\usepackage{xstring}        
\usepackage{etoolbox}       

\usepackage{amsmath}
\usepackage{bm}

\usepackage[dvipsnames, table]{xcolor}

\usepackage{tikz}
    \usetikzlibrary{calc}
    \usetikzlibrary{fadings}
    \usetikzlibrary{backgrounds}
    \usetikzlibrary{external}
        \newenvironment{externalize}{}{}

\usepackage{pgfplots}
    \pgfplotsset{compat=1.16}
    \usepgfplotslibrary{fillbetween}
    \usepackage{pgfmath}
    \usepackage{pgfplotstable}

\usepackage{diagbox}        

\definecolor{lightgreen}{RGB}{146,208,80}
\definecolor{lightblue}{RGB}{0,176,240}
\definecolor{darkgreen}{rgb}{0,0.39,0}
\definecolor{darkblue}{rgb}{0,0,0.5}
\definecolor{lightgreenblue}{RGB}{73,192,160}
\definecolor{darkgreenblue}{rgb}{0,0.195,0.25}

\def\rows{3}
\def\cols{3}
\def\shift{6}
\def\xshift{0}
\def\yshift{0}

\def\scale{0.2}
\def\LineWidthColored{9pt}
\def\lineDist{2.5}

\tikzstyle{node} = [circle, draw, fill=white, inner sep=0pt, text centered, minimum height=3ex*\scale]
\tikzstyle{big} = [circle, draw, fill=white, inner sep=0pt, text centered, minimum height=6ex*\scale]
\tikzstyle{nodec} = [circle, draw, fill=lightgreen, inner sep=0pt, text centered, minimum height=3.5ex*\scale]
\tikzstyle{broken} = [circle, fill=white, densely dashed, inner sep=0pt, text centered, minimum height=3ex*\scale]
\tikzstyle{recphantom} = [nodec, rectangle, minimum height=\lineDist*1ex*\scale, minimum width=\lineDist*1ex*\scale, white]

\tikzstyle{every text node part}=[font=\scriptsize]

\tikzstyle{hamiltonian} = [line width=\scale*\LineWidthColored, rounded corners=2pt, lightgreenblue]
\tikzstyle{edge} = []

\def\brokenQubitsColor{Gray}

\newboolean{drawBrokenQubits}
\setboolean{drawBrokenQubits}{true}

\newboolean{drawQubitsBipartite}
\setboolean{drawQubitsBipartite}{false}
\def\qubitsPartitionColorA{}
\def\qubitsPartitionColorB{black}

\newcommand{\drawGridVertex}[1][node]{%
    \def\RelayedType{{#1}}%
    \drawGridVertexDefaultsRelay
}

\newcommand{\drawGridVertexDefaultsRelay}[3][black]{%
    \ifnumodd{#2 + #3}{
        \node[\RelayedType, #1] at (#2, #3) {};
    }{
        \node[\RelayedType, draw=#1] at (#2, #3) {};
    }
}

\makeatletter
\define@key{chimerakeys}{row}{\def\chimeraRow{#1}}
\define@key{chimerakeys}{col}{\def\chimeraCol{#1}}
\define@key{chimerakeys}{hov}{\def\chimeraHoV{#1}}
\define@key{chimerakeys}{dep}{\def\chimeraDep{#1}}
\tikzdeclarecoordinatesystem{chimera}%
{%
    \setkeys{chimerakeys}{#1}%
    \pgfpointadd{\pgfpointxy{\xshift}{\yshift}}{
    \pgfpointadd{\pgfpointscale{\shift}{\pgfpointadd{\pgfpointxy{\chimeraCol - 1}{1}}{\pgfpointscale{-1}{\pgfpointxy{0}{\chimeraRow}}}}}
                {\ifthenelse{\equal{\chimeraHoV}{h}}{\pgfpointxy{\chimeraDep*1.333 - 1.333}{0}}{\pgfpointxy{2}{3.333 - \chimeraDep*1.333}}}}
}

\newcommand{\chimeraCoord}[4]{(chimera cs:row=#1, col=#2, hov=#3, dep=#4)}

\newcommand{\drawChimera}{

    \foreach \r in {1,...,\rows} {
        \foreach \c in {2,...,\cols} {
            \foreach \d in {1,...,4}{
                \draw[edge] \chimeraCoord{\r}{\c-1}{v}{\d} -- \chimeraCoord{\r}{\c}{v}{\d};
            }
    }}

    \foreach \c in {1,...,\cols} {
        \foreach \r in {2,...,\rows} {
            \foreach \d in {1,...,4}{
                \draw[edge] \chimeraCoord{\r-1}{\c}{h}{\d} -- \chimeraCoord{\r}{\c}{h}{\d};
        }
    }}

    \foreach \r in {1,...,\rows}{
        \foreach \c in {1,...,\cols}{
            \foreach \x in {1,...,4}{
                \foreach \y in {1,...,4}{
                    \draw[edge] \chimeraCoord{\r}{\c}{h}{\x} -- \chimeraCoord{\r}{\c}{v}{\y};
            }}

            \foreach \d in {1,...,4}{
                \node[node] (\r-\c-h\d) at \chimeraCoord{\r}{\c}{h}{\d} {};
                \node[node] (\r-\c-v\d) at \chimeraCoord{\r}{\c}{v}{\d} {};
            }
    }}

}

\newcommand{\clipChimera}[4]{
    \coordinate (m1) at ($(chimera cs:row=#1, col=#2, hov=v, dep=1)!0.5!(chimera cs:row=#1, col=#2, hov=h, dep=1)$);
    \coordinate (m2) at ($(chimera cs:row=#3, col=#4, hov=v, dep=4)!0.5!(chimera cs:row=#3, col=#4, hov=h, dep=4)$);
    \clip ($(m1)+(-2,+2)$) rectangle ($(m2)+(2,-2)$);
}

\newcommand{\drawRectangleInChimera}[5][white]{
    \coordinate (m1) at ($(chimera cs:row=#2, col=#3, hov=v, dep=1)!0.5!(chimera cs:row=#2, col=#3, hov=h, dep=1)$);
    \coordinate (m2) at ($(chimera cs:row=#4, col=#5, hov=v, dep=4)!0.5!(chimera cs:row=#4, col=#5, hov=h, dep=4)$);
    \fill[#1] ($(m1)+(-2,+2)$) rectangle ($(m2)+(2,-2)$);
}

\def\brokenQubits{}

\newcommand{\setBrokenQubits}[1]{
    \def\brokenQubits{}
    \addBrokenQubits{#1}
}

\newcommand{\addBrokenQubits}[1]{
    \foreach \x in #1 {
        \listxadd\brokenQubits{\x}
    }
}

\newcommand{\ifIsBrokenQubit}[3]{
    \xifinlist{#1}{\brokenQubits}{#2}{#3}
}

\newcommand{\ifIsBrokenEdge}[4]{
    \ifboolexpr{ test {\xifinlist{#1}{\brokenQubits}} or test {\xifinlist{#2}{\brokenQubits}} }{#3}{#4}
}

\newcommand{\ifDrawBrokenQubits}[2]{
    \ifthenelse{\boolean{drawBrokenQubits}}{#1}{#2}
}

\newcommand{\ifDrawQubitsBipartite}[2]{
    \ifthenelse{\boolean{drawQubitsBipartite}}{#1}{#2}
}

\newcommand{\drawBrokenChimera}{

    \foreach \r in {1,...,\rows}{
        \foreach \c in {1,...,\cols}{

            \foreach \d in {1,...,4}{
                \drawQubit{\r}{\c}{h}{\d}
                \drawQubit{\r}{\c}{v}{\d}
            }

            \foreach \x in {1,...,4}{
                \foreach \y in {1,...,4}{
                    \drawEdgeIntra{\r}{\c}{\x}{\y}
            }}
    }}

    \foreach \r in {1,...,\rows} {
        \foreach \c in {2,...,\cols} {
            \foreach \x in {1,...,4}{
                \pgfmathparse{int(\c - 1)}
                \drawEdge{\r}{\pgfmathresult}{v}{\x}{\r}{\c}{v}{\x}
            }
    }}

    \foreach \c in {1,...,\cols} {
        \foreach \r in {2,...,\rows} {
            \foreach \x in {1,...,4}{
                \pgfmathparse{int(\r - 1)}
                \drawEdge{\pgfmathresult}{\c}{h}{\x}{\r}{\c}{h}{\x}
            }
    }}
}

\newcommand{\drawQubit}[4]{
    \ifIsBrokenQubit{(#1-#2-#3#4)}{
        \ifDrawBrokenQubits{
            \node[broken, draw=\brokenQubitsColor] (#1-#2-#3#4) at \chimeraCoord{#1}{#2}{#3}{#4} {};
        }{
            \node[node, opacity=0] (#1-#2-#3#4) at \chimeraCoord{#1}{#2}{#3}{#4} {};
        }
    }{
        \ifDrawQubitsBipartite{
            \ifnumodd{\r + \c}{
                \ifthenelse{\isin{v}{#3}}{
                    \node[node, fill=\qubitsPartitionColorA] (#1-#2-#3#4) at \chimeraCoord{#1}{#2}{#3}{#4} {};
                }{
                    \node[node, fill=\qubitsPartitionColorB] (#1-#2-#3#4) at \chimeraCoord{#1}{#2}{#3}{#4} {};
                }
            }{
                \ifthenelse{\isin{v}{#3}}{
                    \node[node, fill=\qubitsPartitionColorB] (#1-#2-#3#4) at \chimeraCoord{#1}{#2}{#3}{#4} {};
                }{
                    \node[node, fill=\qubitsPartitionColorA] (#1-#2-#3#4) at \chimeraCoord{#1}{#2}{#3}{#4} {};
                }
            }
        }{
            \node[node] (#1-#2-#3#4) at \chimeraCoord{#1}{#2}{#3}{#4} {};
        }
    }
}

\newcommand{\drawEdgeIntra}[4]{
    \drawEdge{#1}{#2}{h}{#3}{#1}{#2}{v}{#4}
}

\newcommand{\drawEdge}[8]{
    \ifIsBrokenEdge{(#1-#2-#3#4)}{(#5-#6-#7#8)}{
        \ifDrawBrokenQubits{
            \draw[\brokenQubitsColor, dashed] (#1-#2-#3#4) -- (#5-#6-#7#8);}{}
    }{
        \draw[edge] (#1-#2-#3#4) -- (#5-#6-#7#8);
    }
}

\newcommand{\drawColoredQubit}[2]{
    \node[nodec, fill=#2] at #1 {};
}

\newcommand{\drawColoredEdge}[3]{
    \draw[line width=\scale*\LineWidthColored, #3] #1 -- #2;
}

\newcommand{\drawColoredEdges}[2]{
    \pgfmathsetmacro{\qubitbefore}{}
    \foreach \x in {#1} {
        \drawColoredEdge{\qubitbefore}{\x}{#2}
        \global\let\qubitbefore=\x
    }
    \foreach \x in {#1} {
        \drawColoredQubit{\x}{#2}
    }
}

\newcommand{\drawHorizontalFromTo}[5]{
    \foreach \c in {#3,...,#4} {
        \ifthenelse{\c < #4}{
            \pgfmathtruncatemacro\colBefore{\c + 1}
            \drawColoredEdge{(#1-\colBefore-v#2)}{(#1-\c-v#2)}{#5}
        }{}
        \drawColoredQubit{(#1-\c-v#2)}{#5}
    }
}

\newcommand{\drawVerticalFromTo}[5]{
    \foreach \r in {#3,...,#4} {
        \ifthenelse{\r < #4}{
            \pgfmathtruncatemacro\rowBefore{\r + 1}
            \drawColoredEdge{(\rowBefore-#1-h#2)}{(\r-#1-h#2)}{#5}
        }{}
        \drawColoredQubit{(\r-#1-h#2)}{#5}
    }
}

\newcommand{\drawCrossFromTo}[9]{

    \drawColoredEdge{(#1-#2-v#3)}{(#1-#2-h#4)}{#9}
    \drawHorizontalFromTo{#1}{#3}{#5}{#6}{#9}
    \drawVerticalFromTo{#2}{#4}{#7}{#8}{#9}
}

\newcommand{\drawCrossLow}[5]{
    \drawCrossFromTo{#1}{#2}{#3}{#4}{1}{#1}{#2}{\rows}{#5}
}

\newcommand{\setcolor}[1]{
    \IfEqCase{#1}{%
        {1}{\colorlet{currentcolor}{lightblue}}%
        {2}{\colorlet{currentcolor}{NavyBlue}}%
        {3}{\colorlet{currentcolor}{Blue}}%
        {4}{\colorlet{currentcolor}{darkblue}}%
        {5}{\colorlet{currentcolor}{darkgreen}}%
        {6}{\colorlet{currentcolor}{ForestGreen}}%
        {7}{\colorlet{currentcolor}{lightgreen}}%
        {8}{\colorlet{currentcolor}{GreenYellow}}%
        {9}{\colorlet{currentcolor}{orange}}%
        {10}{\colorlet{currentcolor}{red}}%
        {11}{\colorlet{currentcolor}{purple}}%
        {12}{\colorlet{currentcolor}{violet}}%
    }[\colorlet{currentcolor}{#1}]%
}

\newcounter{colorCounter}
\newcommand{\setcolorCounter}{
    \setcolor{\thecolorCounter}
    \stepcounter{colorCounter}
    \ifthenelse{\thecolorCounter > 12}{\setcounter{colorCounter}{1}}{}
}

\newcommand{\drawCrossLowCC}[4]{
    \setcolorCounter
    \drawCrossLow{#1}{#2}{#3}{#4}{currentcolor}
}

\newcommand{\drawHamiltonian}[3]{
    \draw[#3, rounded corners=0.5pt, very thick] (#1.center) foreach \x in {#2} { -- (\x.center)};
}

\newcommand{\setOppositeDirection}[1]{
    \IfEqCase{#1}{%
        {north}{\def\oppositeDirection{south}}%
        {south}{\def\oppositeDirection{north}}%
        {west}{\def\oppositeDirection{east}}%
        {east}{\def\oppositeDirection{west}}%
    }%
}

\newcommand{\fadeBetween}[5]{
    \coordinate (middle) at ($0.5*#1 + 0.5*#2$);
    \coordinate (start) at ($0.9*#1 + 0.1*#2$);    
    \coordinate (end) at ($0.1*#1 + 0.9*#2$);  
    \setOppositeDirection{#5}
    \drawHamiltonian{start}{{middle}}{path fading=\oppositeDirection, #3!50!#4}
    \drawHamiltonian{end}{{middle}}{path fading=#5, #3!50!#4}
}

\foreach \r in {3, 4, 5} {
    \foreach \c in {1, 2, 3, 6, 7, 8} {
        \foreach \d in {1, 2, 3, 4} {
            \forcsvlist{\listxadd\brokenT}{(\r-\c-h\d), (\r-\c-v\d)}
        }
    }
}

\forcsvlist{\listxadd\brokenTnode}{(1-3-h1), (1-3-h4), (1-3-v2), (1-3-v3),
                                   (1-4-h4), (1-4-v4), (1-4-h3),
                                   (1-5-v1), (1-5-h4),
                                   (1-6-h1), (1-6-h4), (1-6-v2), (1-6-v3),
                                   (2-3-h1), (2-3-h4), (2-3-v2), (2-3-v3),
                                   (2-4-h4), (2-4-v4),
                                   (2-5-h4), (2-5-v1),
                                   (2-6-h1), (2-6-h4), (2-6-v2), (2-6-v3),  
                                   (3-4-h2), (3-4-h3), (3-4-v1), (3-4-v4),
                                   (3-5-h2), (3-5-h3), (3-5-v1), (3-5-v4)}

\forcsvlist{\listxadd\brokenTnodeOdd}{(1-6-h3), (1-6-v4),
                                      (2-3-h2), (2-3-v1),
                                      (3-5-v3), (3-5-h1)}

\foreach \r in {1, 2} {
    \foreach \c in {1, 2, 3} {
        \foreach \d in {1, 2, 3, 4} {
            \forcsvlist{\listxadd\brokenTleft}{(\r-\c-h\d), (\r-\c-v\d)}
        }
    }
}

\foreach \r in {3, 4, 5} {
    \foreach \c in {4, 5} {
        \foreach \d in {1, 2, 3, 4} {
            \forcsvlist{\listxadd\brokenTbelow}{(\r-\c-h\d), (\r-\c-v\d)}
        }
    }
}

\foreach \r in {1, 2} {
    \foreach \c in {3, 4, 5, 6} {
        \foreach \d in {1, 2, 3, 4} {
            \forcsvlist{\listxadd\brokenS}{(\r-\c-h\d), (\r-\c-v\d)}
        }
    }
}
\foreach \r in {5, 6} {
    \foreach \c in {1, 2, 3, 4} {
        \foreach \d in {1, 2, 3, 4} {
            \forcsvlist{\listxadd\brokenS}{(\r-\c-h\d), (\r-\c-v\d)}
        }
    }
}

\begin{document}


\newbox{\orcidboxelli}
\sbox{\orcidboxelli}{\orcidlink{0000-0002-3473-8906}}
\newcommand{\orcidelli}{\usebox{\orcidboxelli}}



\title{Minor Embedding in Broken Chimera and Pegasus Graphs is NP-complete}
\author[1]{Elisabeth Lobe \orcidelli}
\author[1]{Annette Lutz} 
\affil[1]{Institute for Software Technology, German Aerospace Center (DLR), Germany}

\date{\normalsize\today}

\maketitle

\paragraph{Abstract} %
The embedding is an essential step when calculating on the D-Wave machine.  
In this work we show the hardness of the embedding problem for both types of existing hardware,
represented by the Chimera and the Pegasus graphs, containing unavailable qubits. 
We construct certain broken Chimera graphs, where it is hard to find a Hamiltonian cycle. 
As the Hamiltonian cycle problem is a special case of the embedding problem, 
this proves the general complexity result for the Chimera graphs. 
By exploiting the subgraph relation between the Chimera and the Pegasus graphs,  
the proof is then further extended to the Pegasus graphs.

\def\and{, }


    \pause


\section{Introduction} 

\subsection{Background}

Quantum annealing is an emergent technology mainly driven by the developments of the company D-Wave systems. 
Their machines attract a growing number of users from diverse fields by promising to solve their optimization problems faster than classical computers could ever do.
The problem is encoded in a system of quantum bits (\emph{qubits}) with adjustable degrees of pairwise interactions~\cite{dwavedocsQPU}. 
By slowly evolving the quantum system several times, 
the quantum annealers provide a sample of solutions in each run with objective values close to the optimum~\cite{junger2021quantum}. 

The D-Wave machines optimize a so-called \emph{Ising model} with a specific structure, 
which was shown to be NP-complete~\cite{choi2008minor}. 
Thus, in theory, a huge variety of problems can be transferred to them in polynomial time~\cite{lucas2014ising}. 
However, several transformation steps are required to bring the actual problem into the specific Ising format. 
In particular, the limited size of the hardware and the restricted qubit connectivity, which can be represented by a graph, need to be overcome.
We are not aware of any application of practical interest matching the so-called \emph{Chimera} or the newly released \emph{Pegasus} graph directly.
See, e.g., ~\cite{venturelli2015quantum,rieffel2015case,stollenwerk2019quantum,stollenwerk2019flight} for some example applications.

This requires to solve the so-called \emph{embedding} problem, where a set of qubits shall be found for every original vertex to represent it in the hardware.
By applying a strong coupling to the edges between the qubits in such a set, they can act as a single logical vertex
and the union of all neighbors of these qubits yields the desired connectivity~\cite{choi2011minor}. 
We give a more precise definition of an embedding and its relation to the term \emph{minor} in \refSec{embedding}.

\begin{figure}[b]
    \centering
    \def\scale{0.2}
    \begin{externalize}
\begin{tikzpicture}[scale=\scale]

    \drawChimera

    \foreach \rc in {1, 2, 3}{
        \foreach \d in {1, ..., 4}{
            \drawCrossLowCC{\rc}{\rc}{\d}{\d}
        }
    }

\end{tikzpicture}
\end{externalize}
    \caption[Standard triangular complete graph embedding in the ideal Chimera graph]%
            {Standard triangular complete graph embedding in the ideal Chimera graph. 
             Each color represents a single logical vertex in the complete graph}
    \label{fig:complete}
\end{figure}
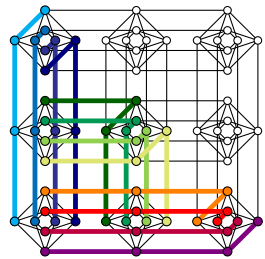

The embedding problem has a straightforward solution for certain well-structured graphs transferred into the ideal Chimera graph.
For example in \refFig{complete}, the standard embedding of the complete graph with 12 vertices is shown, derived from the TRIAD layout introduced in~\cite{choi2011minor}.
It can easily be extended for larger Chimera graphs, where the number of embeddable vertices grows linearly with the size of the Chimera graph. 
This however means, that the number of vertices used in the hardware graph grows quadratically. 
In an ideal Chimera graph in the size of the currently operating \texttt{DW\_2000Q} solver of D-Wave, we can therefore embed only up to 65 completely connected vertices, 
although it contains 2048 qubits~\cite{klymko2014adiabatic}.  

Further easily embeddable graphs can be found for instance in~\cite{lobe2016quadratische}.
In general, those graphs follow the underlying grid structure of the Chimera.
For arbitrary graphs without an obvious relation to such a structure, being too large to simply use the complete graph scheme but small enough that they might fit, 
we always need to find a suitable embedding before we are able to calculate on the machine. 
This still holds for the new hardware architecture with the Pegasus graph, although it yields a higher connectivity~\cite{boothby2020next}. 
 
Above we use the term \emph{ideal} as it refers to the graph provided from the built-in hardware structure
of overlapping superconducting loops forming the qubits, which is for instance described in~\cite{boothby2020next}.   
However, some qubits, or in rare cases also couplings, are switched off by the programming interface of the currently operating hardware, 
because they do not show the expected behaviour~\cite{dwavedocsTopologies}.
As a single such qubit can already break the standard embedding schemes, 
it is not clear anymore how to find an embedding for the complete and other well structured graphs, not to speak of arbitrary graphs. 
Furthermore, after each calibration of the machine, the location of the inaccessible qubits changes. 

Considering the ongoing development of the quantum annealing devices, we expect the number of qubits to grow further.  
But it seems unlikely that a fully connected graph structure can soon be realized due to physical restrictions. 
At the same time, we assume to see a decreasing number of broken qubits compared to the total number, however, they will not vanish completely. 
For calculations on D-Wave's annealing machines, the embedding problem will therefore stay relevant in the long term.

\subsection{Related Work} 

In general, for two arbitrary graphs $G$ and $H$, it is NP-hard to decide whether $G$ can be embedded in $H$.
We go into more detail about this in \refSec{reductions}. 
In their extensive research around such graph relations, Robertson and Seymour could show among other things 
that there exists a polynomial algorithm to decide this question when $G$ is fixed to a single graph and only $H$ is part of the input \cite{robertson1995graph}. 
As it includes the knowledge about so-called forbidden minors of $G$, this result hides constants depending exponentially on the size of $G$. 
A concrete algorithm following Robertson and Seymour's idea is given for hardware graphs with fixed branchwidth in \cite{hicks2004branch}, which is further improved by \cite{adler2011faster}, 
and for planar hardware graphs in \cite{adler2012fast}. 
Although the bounds were improved, the dependence on $G$ is still exponential and thus the algorithm practically infeasible for larger graphs.  

The embedding problem in the quantum annealing context deals with a different case as $G$ is an arbitrary graph derived from the specific application 
while $H$ is fixed to be in the class of all Chimera or Pegasus graphs with broken vertices. 
Even though the hardware graphs are very well structured, no comparable results to Robertson and Seymour's are known in that case. 
In the contrary, we show in this work that the problem remains hard.
This is stated more formally in \refSec{reductions} by \refThm{main}. 

This means in general, the embedding problem is as hard as the actual problem that shall be solved on the D-Wave machine. 
This has some implications in practise. 
Usually, several possibilities to encode an arbitrary optimization problem as an Ising model exist, 
e.g.\ by introducing more vertices, the connectivity between them could be reduced. 
However, in view of our hardness result, we cannot expect to identify an easy to recognize set of graph properties which  
the chosen formulation should fulfil to be better embeddable than others or to be embeddable at all in the current hardware graph. 

Heuristics, as for example presented in~\cite{cai2014practical}, are intended to embed arbitrary graphs 'on-the-fly'. 
Although they work well in practice at the moment, especially for sparse input graphs, with growing hardware sizes, they will most likely not be able to scale as well. 
To overcome such a possible bottleneck, the idea of precalculated templates was already introduced in~\cite{goodrich2018optimizing}.
Virtual hardware graphs form an intermediate layer between the actual hardware and the problem graph and shift the expensive computation away from the user.
For the most general template, the complete graph, an algorithm was found in~\cite{lobe2021embedding} whose runtime is exponential only in the number of broken vertices. 
Other templates are for instance discussed in \cite{serra2019template}.

Our construction to prove the stated complexity result was mainly inspired by the ideas of \cite{itai1982hamilton}. 
There, Itai~et.~al.\ showed that the \textsc{Hamiltonian Cycle Problem} for grid graphs is NP-complete.
As the \textsc{Hamiltonian Cycle Problem} reduces to the embedding problem,   
we adopt several of their definitions and results for the grid graphs and, in the first place, transfer them to the Chimera graphs.  
For this, we introduce the basic concepts in \refSec{preliminaries}. 
Afterwards we show the construction of specific Chimera graphs in \refSec{construction}, 
for which we establish several results about Hamiltonian paths and cycles in \refSec{hamiltonicity}.
With this, we can conclude the proof of the NP-completeness of the \textsc{Hamiltonian Cycle Problem} and thus of the embedding problem for Chimera graphs.
The extension to Pegasus graphs is then shown in \refSec{pegasus}. 
Finally, we give a brief outlook about further research directions in \refSec{outlook}.

    \pause

\section{Preliminaries}\label{sec:preliminaries}

In this section, we introduce the main terms and notations and provide a collection of existing concepts,
recaptured and matched up to give a full picture of the overall proof.
We start with minors and the related embeddings. 
After introducing the specific hardware graph of the D-Wave machine, the Chimera graph, 
we formulate the main question and reduction steps. 
Our construction is based on the ideas of \cite{itai1982hamilton}, whose basic grid definitions are provided at the end of this section. 

In our construction we always refer to simple undirected graphs. 
For the basic graph definitions, we generally follow \cite{korte2008combinatorial}.
For shortness, we identify the tuple $(x,y) \in X \times Y$ for two sets $X, Y$ with the non-commutative product $xy$.
We will also use $vw$ to abbreviate the subset $\{v, w\} \subseteq X$, although $v$ and $w$ commute. 
However, it is clear from the context whether reverting the product ordering is feasible. 
In general, we use $xy$ for integer coordinates in the two-dimensional space, 
while $vw$ refers to an edge of a graph. 

\inputFromHere{preliminaries/Minors}
\pause

\inputFromHere{preliminaries/Chimera}
\pause

\inputFromHere{preliminaries/Reductions}
\pause

\inputFromHere{preliminaries/Grid}

    \pause

\section{Chimera Construction}\label{sec:construction}

In this section, we show the construction of $C(B)$ for an arbitrary $B \in \mathcal{B}$. 
To ensure that the existence of a Hamiltonian cycle is mutually induced between the original graph $B$ and the constructed broken Chimera graph $C(B)$,
we use a very restricted version of the broken Chimera graph. 
Although this might not be necessary for the hardness of the problem, 
meaning most likely we could reduce the number of broken vertices, 
we want to keep the Hamiltonian cycle construction mostly unambitious. 

We start by giving an overview about the overall concept by evolving the underlying grid graph. 
Afterwards, we introduce the representations of the single elements, the vertices and edges. 
Finally, we combine the elements to the full broken Chimera graph corresponding to $B$. 

\inputFromHere{construction/Overview}
\pause

\inputFromHere{construction/Vertices}
\pause

\inputFromHere{construction/Tentacles}
\pause

\inputFromHere{construction/Composition}

    \pause

\section{Hamiltonicity}\label{sec:hamiltonicity}
    
In this section, we establish some results about Hamiltonian paths and cycles, first in the single elements
and finally in the full constructed broken Chimera graph $C(B)$.  
After introducing the overall concept, we evaluate Chimera tentacles first as we build on previous results on Hamiltonian paths in rectangular Chimera graphs there. 
After analysing the vertex elements, we can finally combine the full Hamiltonian cycle. 

\inputFromHere{hamiltonicity/Concept}
\pause

\inputFromHere{hamiltonicity/Edges}
\pause

\inputFromHere{hamiltonicity/Vertices}
\pause

\inputFromHere{hamiltonicity/Assembly}
\pause

\inputFromHere{hamiltonicity/Induction}

    \pause

\section{Transfer to Pegasus}\label{sec:pegasus}

In this section, we briefly want to transfer our findings about the Chimera graph to the newly released hardware structure, the \emph{Pegasus} graph. 
It is derived from the Chimera graph by stretching and shifting the underlying superconducting loops. 
By this, the grid structure of unit cells is preserved but a larger connectivity between the vertices can be realized,
in particular, bridging different unit cells.
As the Pegasus graph is less accessible than the Chimera, we do not go into more details here.  
Please see \cite{boothby2020next} for a complete formal description of the Pegasus graph $P_n$ for $n \in \N$ and its different derived versions.

Here, we concentrate on the 'standard' Pegasus graph, meaning with the default shift, as currently available in hardware by the D-Wave advantage system.
We use the same term 'broken' as for the Chimera graph when considering a vertex-induced subgraph of a Pegasus graph.
Analogously to $\mathcal{C}$, let $\mathcal{P}$ be the set of all finite broken Pegasus graphs.
We can now state a similar question as before, 
where we can show that the hardness of this question relates to the one for the Chimera graph. 

\begin{defi}[\textsc{Broken Pegasus Minor Embedding Problem}]
    Given an arbitrary graph $G$ and some $P \in \mathcal{P}$, is $G$ a minor of $P$, i.e., is $G$ embeddable in $P$?
\end{defi}

\begin{figure}[b!]
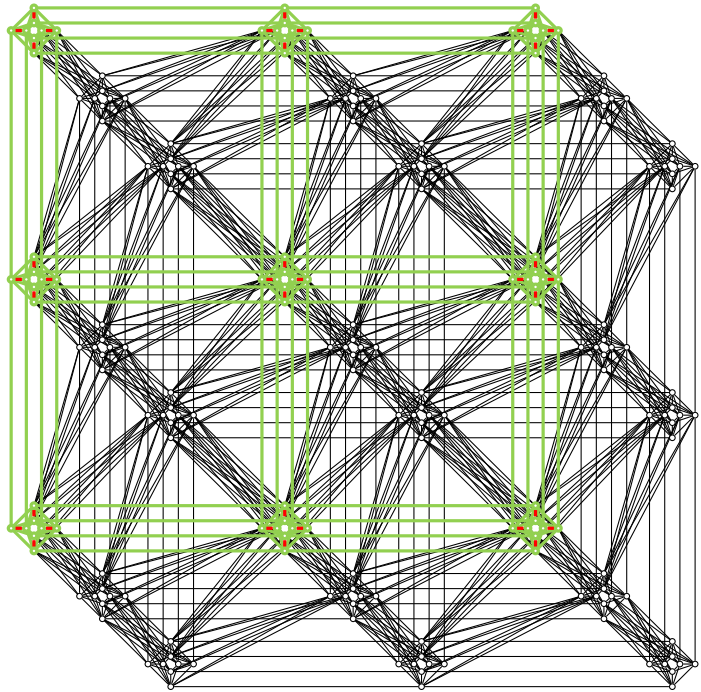

    \centering
    \def\scale{0.15}
    \inputFromHere{../tikz/Pegasus}
    \caption[Inner of the Pegasus graph $P_5$]%
            {Inner of the Pegasus graph $P_5$ 
             without the incomplete unit cells at the boundary}
    \label{fig:pegasus}
\end{figure}

\refFig{pegasus} illustrates the relation between the Chimera and the Pegasus graph.
The highlighted subgraph of the depicted Pegasus graph does nearly have the same connectivity as the Chimera graph 
apart from the additional edges inside the unit cells marked in red.
Let $C^+_{cr}$ be the Chimera graph derived from $C_{cr}$ by adding these edges, thus an edge from the first to the second vertex 
and an edge from the third to the fourth vertex in each partition in each unit cell of $C_{cr}$.
Now we can easily see from the figure, that $C^+_{cr}$ is isomorphic to a vertex-induced subgraph of a (standard) Pegasus graph.
In other words, it is a broken Pegasus graph. 

By enclosing the additional edge between the first and second, respectively, the third and fourth vertex, wherever both vertices are non-broken, 
in each partition of each unit cell in all broken Chimera elements of \refSec{construction}, 
we can construct analogously $\tilde{C}^+(S_v)$ for all $v \in V(B)$ and 
$\tilde{C}^+(T_e)$, respectively, $C^+(T_e)$ for all $e \in E(B)$ and thus $C^+(B)$ for $B \in \mathcal{B}$.
This can still be done in polynomial time and we have $C^+(B) \in \mathcal{P}$.
 
Although $C^+(B)$ is not bipartite anymore, we also keep the chequered pattern of even and odd vertices.
It is easy to see, that all the statements, lemmas and corollaries, in \refSec{hamiltonicity} still hold 
when replacing all occurrences of Chimera elements with the corresponding ones from above. 
In particular, the proof of \refLem{back} remains valid as the additional edges just concern vertices inside a unit cell
and thus the Chimera tentacles are still connected to the vertex elements by only three edges in total.
 
This way, we can deduce the following:

\begin{lem}
    The \textnormal{\textsc{Hamiltonian Cycle Problem}} for graphs $P \in \mathcal{P}$ is NP-complete. 
\end{lem}

Similarly to the treatment of Chimera graphs, this implies

\begin{thm}
    \!The \textnormal{\textsc{Broken Pegasus Minor Embedding Problem}} is NP-complete.
\end{thm}

Note that we could replace the set of edges added between the vertices in each unit cell in $C^+(B)$ by a different one. 
This would not interfere with our Hamiltonian cycle construction nor with the argument for the reverse induction of a Hamiltonian cycle in the original graph.    
Even if we consider a Chimera graph structure but with unit cells forming a complete subgraph, the above argumentation also holds.
Thus, the decisive pattern here is having a vertex-induced subgraph that consists of unit cells with a $K_{4,4}$-subgraph
being arranged and connected in a grid pattern.   
This means, for all new hardware architectures, if they fulfil this property, the embedding problem remains hard in general. 

    \pause

\section{Outlook}\label{sec:outlook}

In the given proof we use a very restricted construction for the broken Chimera graph.
While certain broken vertices are indispensable, for instance those in the odd vertex representation, 
some vertices could be added again and the results would still hold. 
In particular, the 'wholes' between the Chimera tentacles yield a large number of broken vertices:
From \refLem{pppe} we see that the area covered by the broken Chimera graph grows quadratically, 
while the number of unit cells that are actually used in the construction might only depend linearly on the size of original graph. 
Thus, the wholes would occupy a quadratic number of unit cells, 
meaning with growing graph size, the ratio of non-broken vertices would tend to zero.

Currently operating annealing hardware has in turn a very small ratio of broken vertices. 
One possibility to overcome the gap could be to fill up the wholes with non-broken unit cells connected to the tentacles, 
such that they can be covered by return or cross paths, too. 
A first step would be therefore to estimate the best broken vertex ratio which we can achieve with such a construction
within the bounds given by the rectangular representation.
Another open research question is whether we can restrict the broken vertex ratio, for example to a certain linear function in the size of the Chimera graph, 
and still show that the problem is hard.   

Another open problem is if the construction also holds for Pegasus graphs with an arbitrary shift different than the standard one, as explained in \cite{boothby2020next}.  
Can we always find the Chimera graph as a subgraph in these Pegasus graphs like in \refSec{pegasus}? 
As the reduction to the Chimera includes to declare two thirds of the Pegasus vertices to be broken, 
there is also room for further improvement concerning the broken vertex ratio. 

    \pause

    \bibliographystyle{abbrv}
    \bibliography{adds/references}

\begin{thebibliography}{10}

\bibitem{adler2011faster}
I.~Adler, F.~Dorn, F.~V. Fomin, I.~Sau, and D.~M. Thilikos.
\newblock Faster parameterized algorithms for minor containment.
\newblock {\em Theoretical Computer Science}, 412(50):7018--7028, 2011.
\newblock \doi{10.1016/j.tcs.2011.09.015}.

\bibitem{adler2012fast}
I.~Adler, F.~Dorn, F.~V. Fomin, I.~Sau, and D.~M. Thilikos.
\newblock Fast minor testing in planar graphs.
\newblock {\em Algorithmica}, 64(1):69--84, 2012.
\newblock \doi{10.1007/s00453-011-9563-9}.

\bibitem{boothby2020next}
K.~Boothby, P.~Bunyk, J.~Raymond, and A.~Roy.
\newblock Next-generation topology of {{D-Wave}} quantum processors.
\newblock {\em \arxiv{2003.00133}}, 2020.

\bibitem{cai2014practical}
J.~Cai, W.~G. Macready, and A.~Roy.
\newblock A practical heuristic for finding graph minors.
\newblock {\em \arxiv{1406.2741}}, 2014.

\bibitem{choi2008minor}
V.~Choi.
\newblock Minor-embedding in adiabatic quantum computation: {I}. {The}
  parameter setting problem.
\newblock {\em Quantum Information Processing}, 7(5):193--209, 2008.
\newblock \doi{10.1007/s11128-008-0082-9}.

\bibitem{choi2011minor}
V.~Choi.
\newblock Minor-embedding in adiabatic quantum computation: {II}.
  {Minor}-universal graph design.
\newblock {\em Quantum Information Processing}, 10(3):343--353, 2011.
\newblock \doi{10.1007/s11128-010-0200-3}.

\bibitem{dwavedocsTopologies}
{D-Wave Systems Inc.}
\newblock {D-Wave Systems} documentation -- {Getting} started with {D-Wave}
  solvers -- {D-Wave} {QPU} architecture: Topologies.
\newblock
  \href{https://docs.dwavesys.com/docs/latest/c\_gs\_4.html}{https://docs.dwavesys.com/docs/latest/c\_gs\_4.html}.
\newblock visited 2021-10-14.

\bibitem{dwavedocsQPU}
{D-Wave Systems Inc.}
\newblock {D-Wave Systems} documentation -- {QPU} solver datasheet.
\newblock
  \href{https://docs.dwavesys.com/docs/latest/doc\_qpu.html}{https://docs.dwavesys.com/docs/latest/doc\_qpu.html}.
\newblock visited 2021-10-14.

\bibitem{diestel2017graph}
R.~Diestel.
\newblock {\em Graph theory}, volume 173 of {\em Graduate texts in
  mathematics}.
\newblock Springer, 5. edition, 2017.
\newblock \doi{10.1007/978-3-662-53622-3}.

\bibitem{garey1979computers}
M.~R. Garey and D.~S. Johnson.
\newblock {\em Computers and intractability: {A} guide to the theory of
  {NP}-completeness}.
\newblock Series of Books in the Mathematical Sciences. W. H. Freeman and
  Company, 1979.

\bibitem{goodrich2018optimizing}
T.~D. Goodrich, B.~D. Sullivan, and T.~S. Humble.
\newblock Optimizing adiabatic quantum program compilation using a
  graph-theoretic framework.
\newblock {\em Quantum Information Processing}, 17(5):118, 2018.
\newblock \doi{10.1007/s11128-018-1863-4}.

\bibitem{hicks2004branch}
I.~V. Hicks.
\newblock Branch decompositions and minor containment.
\newblock {\em Networks: An International Journal}, 43(1):1--9, 2004.
\newblock \doi{10.1002/net.10099}.

\bibitem{itai1982hamilton}
A.~Itai, C.~H. Papadimitriou, and J.~L. Szwarcfiter.
\newblock Hamilton paths in grid graphs.
\newblock {\em SIAM Journal on Computing}, 11(4):676--686, 1982.
\newblock \doi{10.1137/0211056}.

\bibitem{junger2021quantum}
M.~J{\"u}nger, E.~Lobe, P.~Mutzel, G.~Reinelt, F.~Rendl, G.~Rinaldi, and
  T.~Stollenwerk.
\newblock Quantum annealing versus digital computing: An experimental
  comparison.
\newblock {\em Journal of Experimental Algorithmics (JEA)}, 26:1--30, 2021.
\newblock \doi{10.1145/3459606}.

\bibitem{klymko2014adiabatic}
C.~Klymko, B.~D. Sullivan, and T.~S. Humble.
\newblock Adiabatic quantum programming: minor embedding with hard faults.
\newblock {\em Quantum information processing}, 13(3):709--729, 2014.
\newblock \doi{10.1007/s11128-013-0683-9}.

\bibitem{korte2008combinatorial}
B.~Korte and J.~Vygen.
\newblock {\em Combinatorial optimization: {Theory} and algorithms}, volume~21
  of {\em Algorithms and Combinatorics}.
\newblock Springer, 6. edition, 2018.
\newblock \doi{10.1007/978-3-662-56039-6}.

\bibitem{lobe2016quadratische}
E.~Lobe.
\newblock {Quadratische} bin{\"a}re {Optimierung} ohne {Nebenbedingungen} auf
  {Chimera}-{Graphen} ({German}) [{Quadratic} binary optimization without
  constraints on {Chimera} graphs].
\newblock Master's thesis, Otto-von-Guericke-Universit{\"a}t Magdeburg, 2016.
\newblock \href{https://elib.dlr.de/112063/}{elib.dlr.de/112063/}.

\bibitem{lobe2021embedding}
E.~Lobe, L.~Sch{\"u}rmann, and T.~Stollenwerk.
\newblock Embedding of complete graphs in broken chimera graphs.
\newblock {\em Quantum Information Processing}, 20(7):1--27, 2021.
\newblock \doi{10.1007/s11128-021-03168-z}.

\bibitem{lucas2014ising}
A.~Lucas.
\newblock Ising formulations of many {NP} problems.
\newblock {\em Frontiers in physics}, 2:5, 2014.
\newblock \doi{10.3389/fphy.2014.00005}.

\bibitem{neven2009nips}
H.~Neven, V.~S. Denchev, M.~Drew-Brook, J.~Zhang, W.~G. Macready, and G.~Rose.
\newblock Nips 2009 demonstration: Binary classification using hardware
  implementation of quantum annealing.
\newblock {\em Quantum}, pages 1--17, 2009.

\bibitem{rieffel2015case}
E.~G. Rieffel, D.~Venturelli, B.~O'Gorman, M.~B. Do, E.~M. Prystay, and V.~N.
  Smelyanskiy.
\newblock A case study in programming a quantum annealer for hard operational
  planning problems.
\newblock {\em Quantum Information Processing}, 14(1):1--36, 2015.
\newblock \doi{10.1007/s11128-014-0892-x}.

\bibitem{robertson1995graph}
N.~Robertson and P.~D. Seymour.
\newblock Graph minors. {XIII}. {The} disjoint paths problem.
\newblock {\em Journal of combinatorial theory, Series B}, 63(1):65--110, 1995.
\newblock \doi{10.1006/jctb.1995.1006}.

\bibitem{serra2019template}
T.~Serra, T.~Huang, A.~Raghunathan, and D.~Bergman.
\newblock Template-based minor embedding for adiabatic quantum optimization.
\newblock {\em \arxiv{1910.02179}}, 2019.

\bibitem{stollenwerk2019flight}
T.~Stollenwerk, E.~Lobe, and M.~Jung.
\newblock Flight gate assignment with a quantum annealer.
\newblock In {\em International Workshop on Quantum Technology and Optimization
  Problems}, pages 99--110. Springer, 2019.
\newblock \doi{10.1007/978-3-030-14082-3\_9}.

\bibitem{stollenwerk2019quantum}
T.~Stollenwerk, B.~O'Gorman, D.~Venturelli, S.~Mandr{\`a}, O.~Rodionova, H.~Ng,
  B.~Sridhar, E.~G. Rieffel, and R.~Biswas.
\newblock Quantum annealing applied to de-conflicting optimal trajectories for
  air traffic management.
\newblock {\em IEEE transactions on intelligent transportation systems},
  21(1):285--297, 2019.
\newblock \doi{10.1109/TITS.2019.2891235}.

\bibitem{venturelli2015quantum}
D.~Venturelli, D.~J. Marchand, and G.~Rojo.
\newblock Quantum annealing implementation of job-shop scheduling.
\newblock {\em \arxiv{1506.08479}}, 2015.

\end{thebibliography}


\end{document}